\documentclass[journal=jpccck,manuscript=article]{achemso}

\usepackage{soul}
\usepackage{tabularx}
\usepackage{float}
\usepackage{caption}
\usepackage{subcaption}
\usepackage{tikz}
\usepackage{upgreek}

\usepackage{amsmath}
\usepackage{array}
\usepackage{multirow}
\usepackage{booktabs}

\SectionNumbersOn

\DeclareUnicodeCharacter{2212}{-}
\DeclareUnicodeCharacter{2005}{\hspace{0.25em}}

\author{Ethan Holbrook}
\affiliation{School of Materials Engineering and Birck Nanotechnology Center, Purdue University, West Lafayette, Indiana, 47907 USA}

\author{Matthew P. Kroonblawd}
\affiliation{Physical and Life Sciences Directorate, Lawrence Livermore National Laboratory, Livermore, California 94550, USA}
\email{kroonblawd1@llnl.gov}

\author{H. Keo Springer}
\affiliation{Physical and Life Sciences Directorate, Lawrence Livermore National Laboratory, Livermore, California 94550, USA}

\author{Alejandro Strachan}
\affiliation{School of Materials Engineering and Birck Nanotechnology Center, Purdue University, West Lafayette, Indiana, 47907 USA}
\affiliation{Department of Materials Engineering, Purdue University}
\email{strachan@purdue.edu}

\title{Shock-driven Jetting at HMX Crystal Surface Asperities}

\abbreviations{HMX,TNT,EM,MD,IFE,SFE}

\begin{document}




\begin{abstract}

Shock-induced jetting and recompression are plausible mechanisms for hot spot formation in energetic materials, but remain poorly understood. To this end, we apply all-atom nonreactive molecular dynamics (MD) simulations of a shock-loaded sinusoidal surface asperity in the crystalline explosive HMX (octahydro-1,3,5,7-tetranitro-1,3,5,7-tetrazocine) to quantify the roles of shock strength, asperity geometry, system size, and crystallographic orientation in determining the jetting response. Our simulations span a shock pressure interval of $\approx$4-47~GPa, which captures responses in both detonation initiation and steady-state detonation regimes. We find that shock strength and asperity curvature exert strong influence on jet velocity, temperature, and morphology, while feature size and crystallographic orientation play more modest roles. Jet morphology correlates strongly with the HMX ambient-pressure phase diagram and the narrow temperature range associated with its liquid phase. Coherent jets form when focusing and dissipation are sufficient to drive local temperatures above the melting point, while formation of molecular plumes is seemingly connected to jet temperatures exceeding the liquid-vapor transition temperature. Formation of plumes is unlike the jetting responses seen in metals, which typically exhibit liquid phases that are stable over wider temperature intervals. This work proposes a simple hypothesis for the connection between jet morphology and ambient-pressure phase diagrams that can be readily tested in other explosives and metals. It also provides a foundation for ongoing nonreactive and reactive MD simulations of jet impact and recompression to quantify the role played by jets in forming hot spots in HMX-based explosives. 

\end{abstract}

\clearpage 

\section{Introduction}
\label{sec:Intro}

The physical and chemical responses of materials subject to dynamic shock loading are of broad interest in fields ranging from planetary sciences\cite{Asphaug2006, Ravid2021} and prebiotic chemistry,\cite{chyba_endogenous_1992, Goldman2010, Martins2013, Kroonblawd2019} to materials synthesis and processing,\cite{Thadhani1993} and initiation of energetic materials.\cite{Field1992} Invariably, the materials of interest are microstructurally complex, often exhibiting multiple phases and porosity. Interaction of a shock front with material interfaces, including free surfaces, can concentrate energy into small regions that experience more extreme temperatures, pressures, strain rates, and molecular distortions than in the shock-compressed bulk. These localized regions are important because they can trigger downstream phenomena such as chemical reactions, phase transformations, or material failure. Examples include the initiation of energetic materials,\cite{Bassett2017, Turley2024} mechanically induced chemical transformations in polymers and molecular materials,\cite{Hickenboth2007, Caruso2009, Davis2009, Hamilton2022} and shock- or impact-driven prebiotic chemistry.\cite{Goldman2010, Martins2013, Kroonblawd2019} Shock focusing and localized inelastic deformation at nonplanar free surfaces and interfaces can result in narrow, high-velocity material flow. Richtmyer-Meshkov instability\cite{Richtmyer1960, Meshkov1969, Brouillette2002} provides a related framework for understanding perturbation growth through local baroclinic torques that develop when a shock accelerates a density-mismatched interface. Jet formation depends on shock strength, material phase diagram, feature size and aspect ratio, and impedance contrast. While the conditions that lead to jetting are generally understood for fluids and simple solids such as cubic metallic crystals, these responses are poorly characterized for low-symmetry molecular crystals that are typical of most organic energetic materials.

In the case of energetic materials, jetting and subsequent recompression of the material dominate the collapse of porosity under strong shock conditions (e.g., $P_{\rm{shock}} > 10$~GPa),\cite{Holian2002, Wood2015, Li2022} which can lead to formation of hot spots that ignite chemical reactions.\cite{Bassett2017, Turley2024} Experiments have revealed the importance of pore collapse in shock initiation sensitivity\cite{Dattelbaum2010, Duarte2020} and this knowledge has been used to design switchable explosives.\cite{Brown2023,Lawrence2024} Engineering defects to construct desirable detonation wave structures is also becoming increasingly viable.\cite{Wilkening2025} However, the molecular and nanoscale phenomena underlying jetting and pore collapse are difficult to observe experimentally due to the small length scales, fast timescales, and extreme conditions involved.

Experimental diagnostic challenges motivated early applications of continuum mechanics computer simulations to better understand hot spot formation mechanisms in the 1960s.\cite{mader_1965} Modern continuum simulations adopt complex mesh-based integration schemes, which are coupled to increasingly sophisticated multiphysics material models to more accurately capture dynamic responses.\cite{Menikoff2004, Austin2015, Wood2018, Das2021, Duarte2021, Herrin2025, roy2026} Continuum simulations can capture the overall hydrodynamics of the jetting process and have revealed sensitivity to the transition from viscoplastic to hydrodynamic behavior and the strong dependence of hot spot formation on void shape characteristics,\cite{Rai2017, Springer2018, Rai2018, Wood2018} including for highly curved geometries that induce jetting.\cite{springer_2023} Fidelity of these simulations is gated at both ends by computational mesh resolution. Too coarse of a mesh results in overpredicting hot spot mass\cite{Menikoff2004} while refinement below the validity of the continuum constitutive description produces artificial localization.\cite{Herrin2025} In both limits, difficulties and uncertainty arise because continuum material models necessarily approximate molecular-level processes through constitutive relations, including localized crystal plasticity, melting, and vaporization. Size effects and suitability of continuum descriptions become important considerations as the defects controlling hot spot formation shrink below a micron.\cite{Wood2018} Energetic materials often exhibit complicated void defect size distributions with substantial population that is well below this size.\cite{LeePerry2018}

All-atom molecular dynamics (MD) simulations can resolve hot spot formation and jetting with molecular resolution, while using a unified description of coupling between elastic and inelastic mechanical deformations, phase transformations, transport, and chemistry. Recent advances in GPU-accelerated high-performance computing have enabled MD simulations of organic energetic materials with multimicron-scale computational domains and defect sizes spanning hundreds of nanometers.\cite{kober_2024, Kroonblawd2025, yen_2026} All-atom studies have shown that temperature and pressure alone do not fully characterize the state of shock-collapsed pores,\cite{Wood2015, betterhalf2021} that continuum heat transport theory holds on the nanoscale,\cite{Kroonblawd2021} and that interfaces in multi-component systems can induce asymmetries in hot spot formation\cite{kroonblawd_2026} and accelerate chemical reactivity.\cite{Macatangay2024} Direct comparisons between MD and continuum models have also identified a complicated interplay between mechanical strength and phase transitions in mediating plasticity leading to hot spot formation.\cite{Das2021} Pore shape, and specifically pore curvature at the upstream surface, governs the jetting expansion of the material into the pore. Cylindrical pores are most commonly studied, but pore surface orientation, curvature, and length all impact the resulting temperature and pressure states.\cite{Li2020, Li2022, okafor_2025} Despite this progress, the conditions needed to form shock-driven jets in molecular crystals are not well understood.

Because jetting has been studied more extensively in metals than in molecular energetic materials, metallic systems provide a useful reference for assessing the roles of geometry, shock strength, and crystal orientation. Multiple MD studies have focused on the effects of surface roughness, shock strength, and crystal orientation on jetting.\cite{Germann2009, Li2014, Cherne2015, Durand2017, Dyachkov2017, Pereira2023, Hamilton2025} Sinusoidal surface asperities are a common geometry of focus, since they can be summarized with only a few parameters including the amplitude $h_0$ and the wavenumber $k$, which is related to the wavelength $\lambda$ as $k = {2\pi}/{\lambda}$. Continuum simulations indicate that aspect ratio $kh_0$ controls the initial growth of the jet, with $kh_0 = 1$ being at the edge of a linear response regime and where values above unity lead to increasingly nonlinear responses.\cite{Dimonte2010} Multiple MD studies on metals show that shocked surface defects can produce liquid-like jets with a stem-and-bulb morphology,\cite{Germann2009, Li2014, Cherne2015, Dyachkov2017, Hamilton2025} although jet breakup and arcing can also occur depending on the underlying crystal orientation and asperity geometry.\cite{Li2014, Durand2017} Near-surface semicircular pores can also lead to jets, but increasing pore depth can effectively attenuate jets before they break the surface.\cite{Hamilton2025}

We focus here on identifying the geometric and thermodynamic factors that lead to jetting in a prototypical organic energetic material. To this end, we use all-atom MD simulations to probe shock interactions with a sinusoidal surface asperity in oriented crystals of the explosive HMX (octahydro-1,3,5,7-tetranitro-1,3,5,7-tetrazocine). HMX is used in a wide range of explosive formulations, including in composites with polymers and with melt-stable explosives such as TNT (trinitrotoluene).\cite{gibbs_1980} This study aims to quantify how shock strength, surface geometry, and crystallographic orientation jointly determine whether an HMX surface asperity produces a coherent jet. The downstream impact of these jets on hot spot formation is the subject of ongoing work. Through MD simulations, we identify distinct jetting regimes and relate their evolution to shock-induced heating and the HMX phase diagram, particularly the limited stability range of the liquid phase. Correlations between jetting response and the ambient-pressure phase diagram identified here suggest a testable hypothesis for analogous jet morphological evolution in other shock-loaded materials.

\section{Methods}
\label{sec:Methods}

\subsection{MD Details} 
\label{sec:MethodsMD}

Classical MD simulations were performed using the LAMMPS simulation software\cite{Thompson2022} with Kokkos GPU acceleration\cite{Trott2022} and the widely used all-atom nonreactive force field (FF) for HMX originally developed by Smith and Bharadwaj.\cite{Smith1999} All simulations were performed using 3D periodic simulation cells (described below) with nonbonded interactions evaluated in real space using an 11~{\AA} cutoff and the Wolf potential\cite{Wolf1999} for electrostatic interactions. Implementation details and example LAMMPS input files are given in previous work.\cite{Holbrook2025} Trajectories were integrated using a 0.5~fs timestep and a short-range repulsive term (as used previously\cite{Holbrook2025}) unless otherwise noted. Initial thermal configurations for shock simulations were obtained using constant-temperature ($NVT$) equations of motion with a Langevin thermostat. \cite{Brunger1984,DUNWEG1991} Shock simulations were performed using a reverse-ballistic configuration\cite{Holian1998} with constant-energy ($NVE$) equations of motion. In these simulations, a flexible sample impacts a rigid 2~nm thick piston (three-four crystal layers) with net velocity $-u_{\rm{p}}\hat{\mathbf{z}}$, where $u_{\rm{p}}$ is the intended particle velocity. Calculation stability of systems undergoing significant mass transfer can be sensitive to domain decomposition details; we configured the domains to fully span the cell along the shock direction $z$ using the LAMMPS command ``processors * * 1'' to limit excessive atom migration across processors.

\subsection{Model Systems}
\label{sec:MethodsModelSystems}

\begin{figure}[t!]
  \centering
  \includegraphics[width=0.98\textwidth]{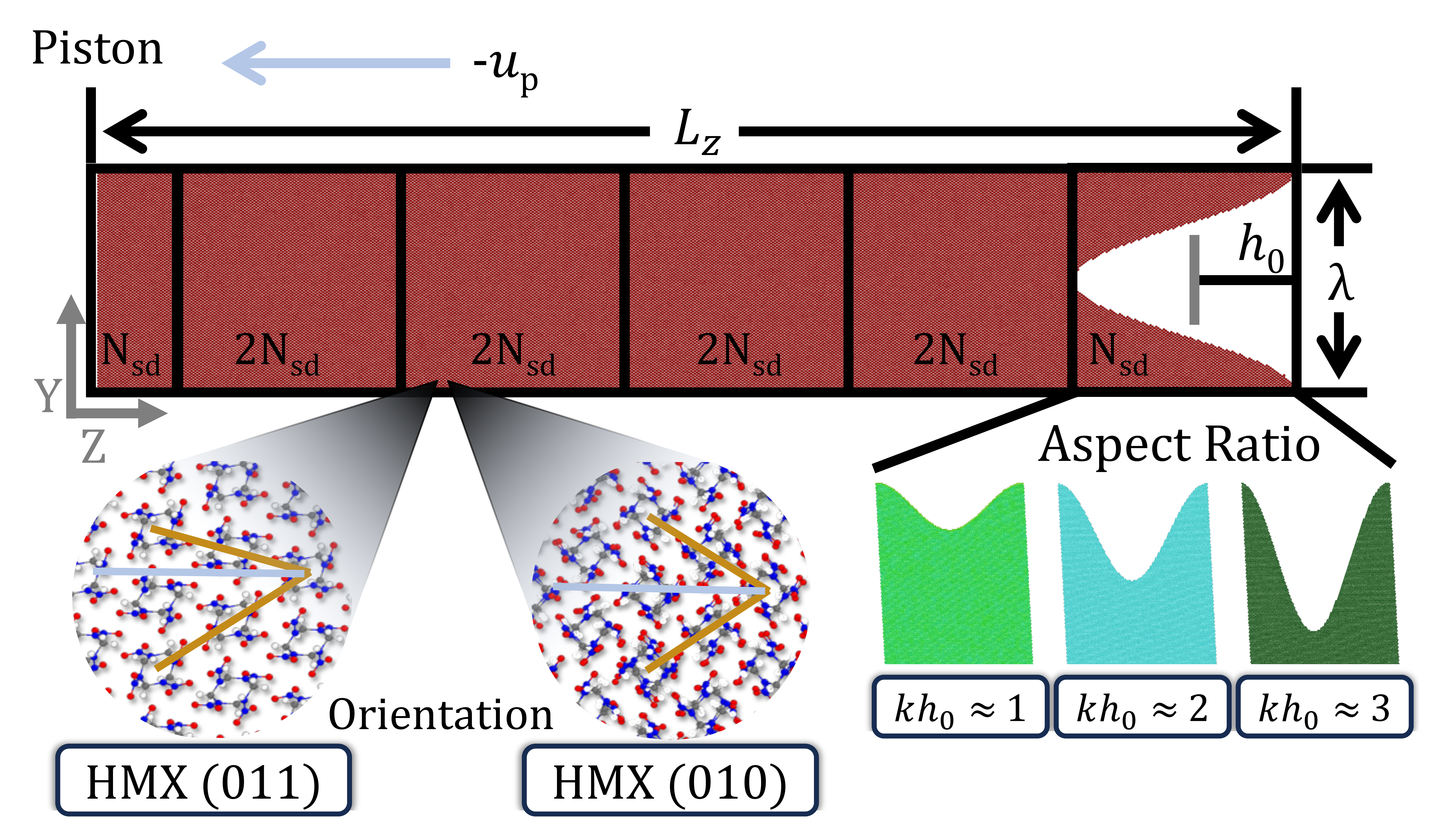}
  \caption{Schematic system constructions with a sinusoidal surface asperity.}
  \label{fig:system_cases}
\end{figure}

Shock-induced jetting at sinusoidal surface asperities was studied using quasi-2D simulation cells containing oriented $\beta$-HMX single crystals that were constructed following the schematic given in Figure~\ref{fig:system_cases}. Two shock directions were considered that were oriented normal to the (010) and (011) facets in the $P2_{1}/n$ setting. These two facets were chosen because they are among the most morphologically important facets,\cite{Dick2004, Gallagher2015, Kroonblawd2016} with (011) being the lowest-energy facet,\cite{Holbrook2025} and because they exhibit distinctly different symmetries about the shock axis. All constructions were prepared using the Generalized Crystal-Cutting Method\cite{Kroonblawd2016} (GCCM) with equilibrium lattice parameters determined\cite{Holbrook2025} at (300~K, 1~atm) following the approach summarized below and described in detail in the SI. Cells were oriented such that $\mathbf{a} = [100]$ was exactly along $x$ and the chosen shock direction was exactly along $z$. The quasi-2D geometry treats the $x$ direction as thin ($L_{x} \approx 4$~nm) and the $y$ and $z$ dimensions as extended, with $L_{y}$ and $L_{z}$ set based on the asperity geometry. All dimensions are more than twice the FF cutoff distance.

The surface asperity geometry was specified by a cut function 
\begin{equation}
z(y) = h_{0} \cos(ky) + z_{0},
\label{eq:cutfunction}
\end{equation}
which describes the sinusoid in terms of its amplitude $h_{0}$ and wavenumber $k$, which is related to the asperity wavelength $\lambda$ as 
\begin{equation}
k = \frac{2\pi}{\lambda}.
\label{eq:wavenumber}
\end{equation}
Here, $z_{0}$ controls the off-set for the cut function and is chosen so that the peak of the asperity is set at $L_{z} = 11h_{0}$, the total crystal length in the shock direction. The cut function phase is also arbitrary given the simulation cell periodicity in $y$, since by definition $L_{y} = \lambda$. From Equation~\ref{eq:wavenumber}, it can be seen that the quantity $kh_{0}$ specifies the asperity height-to-wavelength aspect ratio (i.e., $kh_{0} \propto h_{0}/\lambda$). The surface asperity region dimensions are $2h_{0} \times \lambda$ and contain $N_{\rm{sd}}$ molecules.

\begin{table}[t!]
\centering
\caption{Geometric parameters for simulation cell constructions}
\begin{tabular}{lccclcc}
\hline \hline
Case Name 		& 
Orientation 	& 
$\lambda$ (nm) 	& 
$h_0$ (nm) 		&
$kh_0$ 			& 
$L_{x}$ (nm)	& 
$L_{z}$ (nm) 	\\
\hline
(010)-30nm 	& (010) & 30.4	& 15.2 & $3.11$   & 3.29 & 164 \\
(010)-60nm  & (010) & 60.8	& 30.4 & $3.11$   & 3.29 & 329 \\
(011)-$3$ 	& (011) & 77.2	& 38.6 & $3.11$   & 4.59 & 346 \\
(011)-$2$	& (011) & 77.2	& 25.0 & $2.04$   & 4.59 & 200 \\
(011)-$1$ 	& (011) & 77.2	& 12.5 & $1.00$   & 4.59 & 293 \\
\hline \hline
\end{tabular}
\label{tab:geometry}
\end{table}

Five distinct simulation cell geometries were considered, which are listed in Table~\ref{tab:geometry}. The first two cases were designed to examine asperity wavelength effects and share the (010) crystallographic orientation, aspect ratio $kh_{0} \approx 3$, and quasi-2D thickness $L_{x}$, but vary the wavelength $\lambda$ and cell length $L_{z}$ by a factor of two. The last three cases consider a different crystal orientation, namely (011), and specifically vary the aspect ratio $kh_{0}$. The third case, (011)-3, was designed to correspond closely to the (010)-60nm case, albeit with a modestly larger wavelength $\lambda$. The (010) and (011) cells were constructed with slightly different protocols because the latter were designed with the GCCM lattice-matching algorithm to be commensurate with a down-stream crystalline target. Jet impact will be examined in future work.

The (010) cells were all based on replications of a non-primitive supercell that was identified by the single-crystal version of the GCCM code and had cell vectors $\mathbf{x}_{1} = \mathbf{a}$, $\mathbf{x}_{2} = -7\mathbf{a} - 40\mathbf{c}$, and  $\mathbf{x}_{3} = \mathbf{b}$. This yielded a pseudo-orthorhombic cell with angles $90.00^{\rm{o}} \pm 0.03^{\rm{o}}$. A timestep of 0.1~fs was used with no short-range repulsive term (active under 1.2~{\AA}) for all (010) simulation runs. Thermal configurations were obtained from a 50~ps $NVT$ trajectory integrated with a Langevin thermostat with time constant 0.1~ps. Residual stresses can develop due to free surface and fixed transverse dimensions. Nonzero stress components were compressive and under 10~MPa, which is more than 400 times smaller than the weakest shock pressure considered in this work.

The (011) cells were constructed from replications of a non-primitive supercell that was identified by the crystal-crystal interface version of the GCCM code and had cell vectors $\mathbf{x}_{1} = 7\mathbf{a}$, $\mathbf{x}_{2} = 10\mathbf{b} - 10\mathbf{c}$, and  $\mathbf{x}_{3} = 7\mathbf{a} + 9\mathbf{b} + 19\mathbf{c}$. This (011) non-primitive supercell was identified to match with the (001) facet of TNT crystal and also had a small imposed strain to ensure that the two crystals could be matched in the same periodic cell. Lattice matching necessitated larger deviations from an orthorhombic cell ($90^{\rm{o}} \pm 8^{\rm{o}}$) and also required a somewhat larger $L_{x}$. The imposed volumetric and deviatoric strains were below 1\% and the individual strain components are listed in the SI. Both the free surface and imposed strain can result in residual stresses. Substantially longer $NVT$ trajectories (0.5-1.5~ns) with a weaker thermostat time constant (1.0~ps) were necessary to efficiently relax these stresses to a steady state (shown in the SI). The residual compressive stresses were somewhat larger, being roughly 50~MPa, 75~MPa, and $<$10~MPa in the $x$, $y$, and $z$ dimensions, but these are still 60 times smaller than the weakest shock pressure considered.

\subsection{Trajectory Analysis}
\label{sec:MethodsAnalysis}

Trajectories were analyzed using well-established techniques that are described in detail in previous work.\cite{Kroonblawd2022_MSMSE} Briefly, this analysis used atomic positions, velocities, and stress components to compute molecular center of mass velocity, temperature, stress tensor, and strain tensor. An additional local averaging step was applied by taking the arithmetic average over all molecules whose centers of mass were within a 1.5~nm radius sphere about a given molecule. Molecular temperatures were calculated from the kinetic energy with the rigid translational motion removed. Stresses include both the kinetic and configurational virial contributions and were normalized by the local average sphere volume. Molecular strain tensors were computed using OVITO\cite{ovito} with the preshock configuration as a reference state. Local scalar volumetric and deviatoric strains were determined using the standard $J_{1}$ and $J_{2}$ rotational invariants. The shock pressure was taken to be the $P_{zz}$ component of the stress tensor with the understanding that the post-shock state is in general nonhydrostatic. All system configuration snapshots were rendered with OVITO.\cite{ovito}

Shock pressure and velocity $u_{\rm{s}}$ were measured by tracking the shock front using $z$-$t$ diagrams of $P_{zz}$. The simulation cell was binned along the shock direction $z$ with a fixed grid ($\Delta z = \sim2.5$~nm) and the average $P_{zz}$ was determined over all molecules in each bin every 1~ps. The shocked region behind the front was identified as the set of contiguous bins that met two criteria: (1) the local $P_{zz}$ exceeded 90\% of the profile maximum; and (2) the local gradient fell below a tolerance of $2 \times 10^{-4}$ GPa/{\AA}, with a minimum plateau width of 5 bins (12.5~nm). The shock pressure was extracted as the mean $P_{zz}$ over the plateau bins. The shock velocity in the piston frame $u_{\rm{w}}$ was determined from the slope of the shock front position in a $z$-$t$ diagram. Note that the shock travels at velocity $u_{\rm{w}} = u_{\rm{s}} - u_{\rm{p}}$ in the frame fixed on the piston, where $u_{\rm{p}}$ is the initial negative net velocity imposed on the flexible sample. Example $z$-$t$ diagrams with shock front identification are included for the 60~nm systems in the SI.

\section{Results and Discussion}
\label{sec:ResultsandDiscussion}

\subsection{Shock Strength Dependence}
\label{sec:ResultsUp}

Surface asperity jetting response was predicted for six distinct shock strengths ranging from particle velocity $u_{\rm{p}} = 0.5$~km/s to 3.0~km/s using (010)-oriented HMX cells with two different asperity wavelengths ($\lambda = 30$~nm and 60~nm) and with aspect ratio $kh_{0} = 3$. We focus here on general characteristics of shock interactions with a sinusoidal surface asperity, reserving detailed inspections of size, crystal orientation, and asperity geometry effects for subsequent sections. Since $kh_{0} = 3$, the behavior is expected to be firmly in the nonlinear/jetting regime for shocked sinusoidal interfaces.\cite{Dimonte2010} For reference, HMX-based explosives typically exhibit a shock initiation threshold pressure near 1-2~GPa\cite{Dick1998,Handley2018}, exhibit a $\beta \rightarrow \phi$ solid-solid phase transition at $P_{\phi} = 27$~GPa\cite{Hare2004,Zaug2018}, have Chapman-Jouguet pressure $P_{\rm{CJ}} \approx 35$~GPa \cite{Fritz1996,Menikoff2006}, and von Neumann spike pressure $P_{\rm{VN}} \approx 57$~GPa\cite{Menikoff2006}. Studies on detonation in HMX single crystals are scarce, but show higher pressures than typical HMX-based explosive formulations, with values of $P_{\rm{CJ}} \approx 43$~GPa and $P_{\rm{VN}} \approx 63$~GPa.\cite{Fedorov2011}

\begin{figure}[p!]
  \centering
  \includegraphics[width=0.47\textwidth]{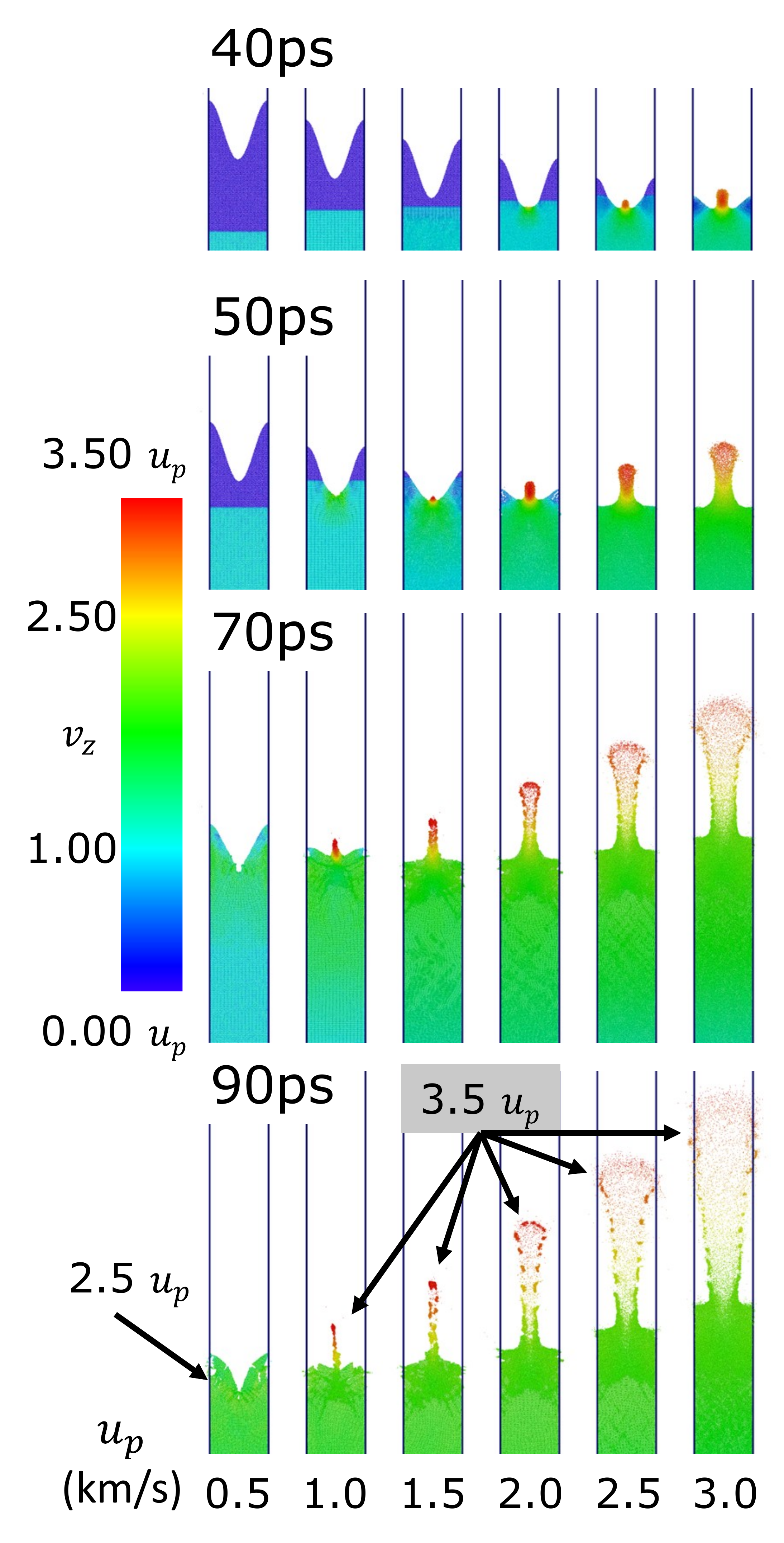}
  \caption{Simulation snapshots showing time sequences of the (010)-60nm, $kh_{0} \approx 3$ configuration at six different shock strengths. Molecules are colored by velocity component $v_{z}$ in the frame of the moving shock, which has expected values of $1.0u_{\rm{p}}$ in the shocked bulk far from free surfaces and $2.0u_{\rm{p}}$ at a planar surface under free expansion. Annotations denote the fastest moving molecules in each case, which reach roughly $2.5u_{\rm{p}}$ in the $u_{\rm{p}} = 0.5$~km/s case and approximately $3.5u_{\rm{p}}$ in cases where $u_{\rm{p}} \geq 1.0$~km/s.}
  \label{fig:Up}
\end{figure}

Snapshot sequences showing the qualitative differences between these cases are given in Figure~\ref{fig:Up} and corresponding shock state variables are listed in Table~\ref{tab:30nm-60nm-UsUp}. We measure and report the shock speed $u_{\rm{s}}$, the shock pressure $P_{zz}$, and the peak velocity of surface molecules $v_{z}^{99.9\%}$, taken as the 99.9\% percentile of the molecular $v_{z}$ distribution. Molecular velocity is measured in the frame where the unshocked material is at rest. Based on the measured shock pressures, cases with $u_{\rm{p}} \leq 2.0$~km/s correspond to the typical range for shock initiation, $u_{\rm{p}} = 2.0$~km/s corresponds to $P_{\phi}$, 2.5~km/s corresponds to the lower range of $P_{\rm{CJ}}$, and 3.0~km/s is above $P_{\rm{CJ}}$. The 3.0~km/s case approaches the range of $P_{\rm{VN}}$, but the peak spike pressure is is not fully captured in our $u_{\rm{p}}$ assay. It is immediately apparent that the jet morphology evolves significantly across this $u_{\rm{p}}$ range.

\begin{table}[t!]
\centering
\small
\setlength{\tabcolsep}{4pt}
\caption{$kh_{0} \approx 3$ shock conditions for the HMX $(010)$-30nm and $(010)$-60nm cases.}
\label{tab:30nm-60nm-UsUp}
\begin{tabular}{cccc@{\hspace{5em}}cccc}
\hline \hline
\multicolumn{4}{c}{HMX $(010)$-30nm } &
\multicolumn{4}{c}{HMX $(010)$-60nm } \\
\hline
$u_{\rm{p}}$ & $u_{\rm{s}}$ & $P_{zz}$ & $v_{z}^{99.9\%} / u_{\rm{p}}$ &
$u_{\rm{p}}$ & $u_{\rm{s}}$ & $P_{zz}$ & $v_{z}^{99.9\%} / u_{\rm{p}}$ \\
(km/s) & (km/s) & (GPa) & (unitless) &
(km/s) & (km/s) & (GPa) & (unitless) \\
\hline
0.5 & 4.76 & $4.44 \pm 0.09$  & 2.44 &
0.5 & 4.76 & $4.51 \pm 0.03$  & 2.60 \\
1.0 & 5.91 & $10.91 \pm 0.03$ & 3.32 &
1.0 & 5.86 & $10.92 \pm 0.03$ & 3.60 \\
1.5 & 6.41 & $17.85 \pm 0.02$ & 3.32 &
1.5 & 6.34 & $17.80 \pm 0.04$ & 3.49 \\
2.0 & 7.09 & $26.38 \pm 0.11$ & 3.39 &
2.0 & 7.00 & $26.39 \pm 0.09$ & 3.46 \\
2.5 & 7.76 & $36.02 \pm 0.05$ & 3.33 &
2.5 & 7.66 & $36.01 \pm 0.10$ & 3.40 \\
3.0 & 8.49 & $46.84 \pm 0.12$ & 3.29 &
3.0 & 8.35 & $46.86 \pm 0.05$ & 3.35 \\
\hline \hline
\end{tabular}
\end{table}

\begin{figure}[t!]
  \centering
  \includegraphics[width=0.98\textwidth]{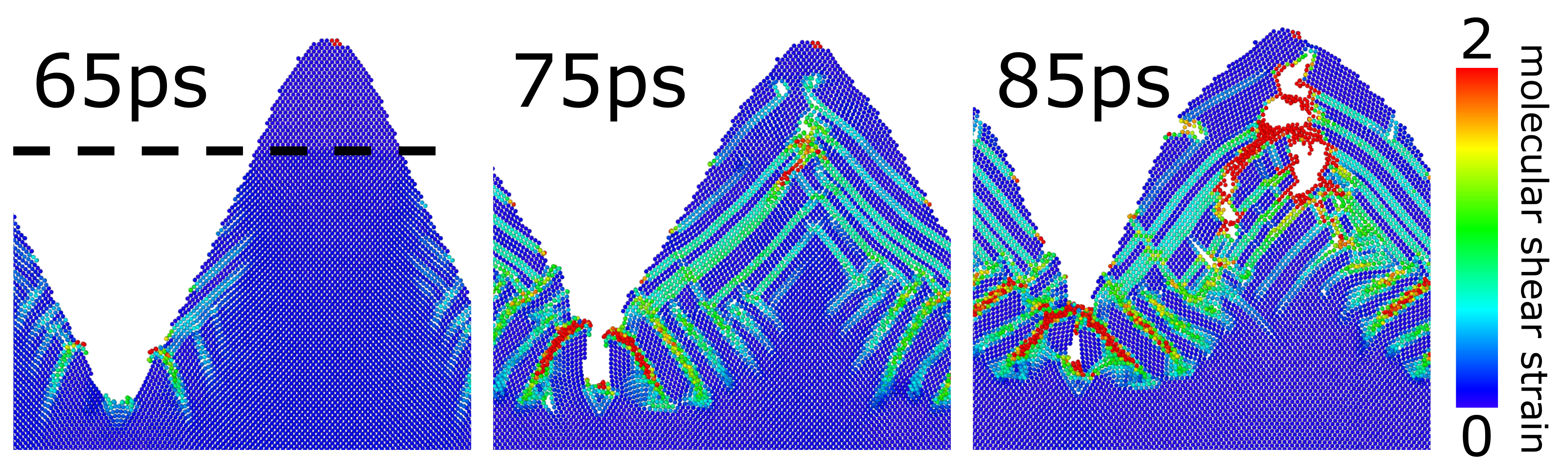}
  \caption{Simulation snapshots showing the time evolution of plastic deformation at the surface asperity in the (010)-60nm case with $kh_{0} = 3$ and $u_{\rm{p}} = 0.5$~km/s. Molecules are colored according to local shear strain and each panel depicts a partially replicated cell showing a domain that is 1.1$\times$ the asperity period.}
  \label{fig:shear_strain}
\end{figure}

At $u_{\rm{p}} = 0.5$~km/s, plastic deformation and damage nucleate near the surface but these deformations do not lead to ejecta, unlike higher $u_{\rm{p}}$ cases which exhibit jet formation. A detailed inspection of the plastic deformation in the $u_{\rm{p}} = 0.5$~km/s case is given in Figure~\ref{fig:shear_strain}, which shows that shear bands nucleate at the free surface as the shock sweeps across the surface asperity (see 65~ps snapshot). This plastic deformation extrudes material into the surface trough and propagates back into the crystal. Shear bands eventually propagate across the entire asperity by 75~ps and the rarefaction wave caused by the interaction of the shock with the free surface results in local damage. Despite this incipient spall, no ejecta are observed (see 85~ps snapshot). Cross inspection with Figure~\ref{fig:Up} shows that the material slightly under the surface (at the tip of the arrow) reaches a velocity of approximately $2.5u_{\rm{p}}$, which is greater than the value of $2.0u_{\rm{p}}$ expected for a flat surface\cite{eliezer2021free}. Plastic deformation is not specific to this case and is present throughout the examined systems. However, as shock strength increases, jet formation and hydrodynamic flow become the dominant deformation modes, reducing the relative role of crystal plasticity.

At $u_{\rm{p}} = 1.0$~km/s, shock focusing around the trough of the asperity results in jetting with a peak material velocity reaching $3.6u_{\rm{p}}$ in the (010)-60nm case. As it happens, this velocity multiplier for the ejecta is the highest for this shock strength and decreases for higher $u_{\rm{p}}$. We note that a maximum in the particle velocity amplification by the asperity is less clear in the smaller (010)-30nm case; instead, the ratios $v_{z}^{99.9\%} / u_{\rm{p}}$ capturing the relative peak ejecta velocity are all equivalent within $0.1u_{\rm{p}}$.

At $u_{\rm{p}} = 1.5$~km/s and above, increasingly wider jets form that ultimately break up into clusters of ejecta. With the larger (010)-60nm simulations, the peak $v_{\rm{z}}^{99.9\%}$ of the ejecta steadily increases with shock pressure, but its value relative to $u_{\rm{p}}$ declines with shock pressure. A distinct change in the morphology of the jet occurs for $u_{\rm{p}} \geq 2.0$~km/s, wherein the central core of the jet clearly vaporizes, resulting in lateral expansion of the jet with a discontinuous liquid outer shell that exhibits a break-up similar to the $u_{\rm{p}} = 1.5$~km/s case. Morphology of the $u_{\rm{p}} = 2.5$~km/s and 3.0~km/s cases maintains a very similar structure as the $u_{\rm{p}} = 2.0$~km/s case, with the exception that the lateral expansion of the plume generally increases with $u_{\rm{p}}$.

\begin{figure}[t!]
  \centering
  \includegraphics[width=0.98\textwidth]{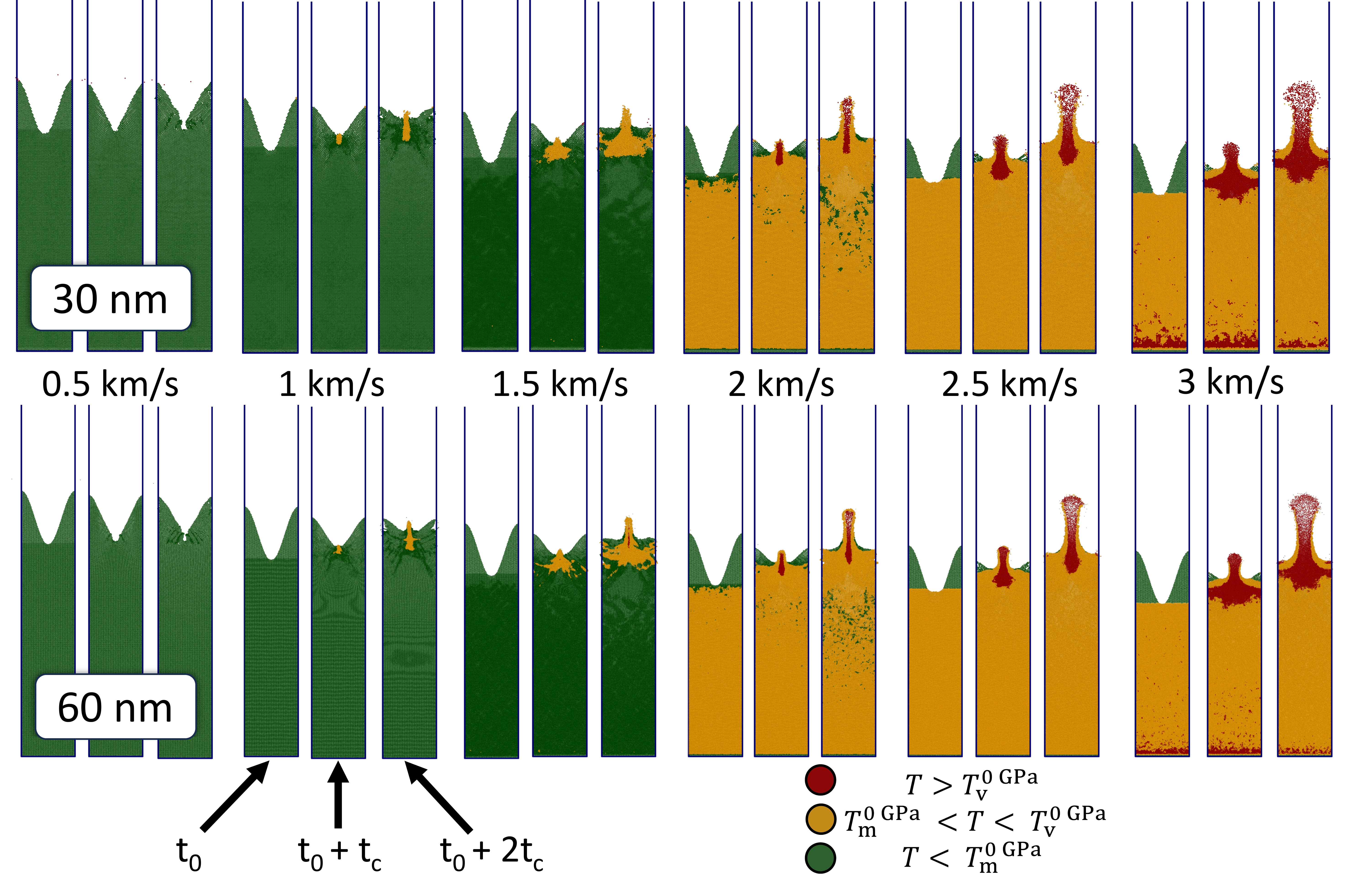}
  \caption{Simulation snapshots showing local temperatures in the (010)-30nm and (010)-60nm cases with $kh_{0} = 3$ at six different shock strengths. Molecules are colored green, yellow, and red to denote temperatures that respectively correspond to the solid, liquid, and vapor phases at $P = 1$~atm. Each grouping of three corresponds to a given simulation size and $u_{\rm{p}}$ with snapshots taken at times $t_{0}$, $t_{0} + t_{\rm{c}}$, and $t_{0} + 2t_{\rm{c}}$, where $t_{0}$ is the shock arrival time at the trough of the asperity and $t_{\rm{c}}$ is the time required for the shock to transit from the trough to the crest of the asperity.}
  \label{fig:size}
\end{figure}

Having examined the local velocity response associated with the range of input shock pressures, we now focus on a qualitative inspection of temperature evolution and phase change. Figure~\ref{fig:size} shows snapshots of each simulation with color denoting local temperature ranges taken at three distinct times. Higher resolution snapshots are given in the SI. Specifically, we categorize molecules based on instantaneous kinetic temperature using the MD-predicted HMX solid-liquid and liquid-vapor phase boundaries at $P = 1$~atm, which previous work shows to be respectively $T_{\rm{m}}^{\rm{0 GPa}} = 578$~K and approximately $T_{\rm{v}}^{\rm{0 GPa}} = 1200$~K.\cite{tow2022, Kroonblawd2023} Note that $P \rightarrow 0$~GPa for material undergoing free expansion, as is the case here. Based on a temperature definition, the color scheme distinguishes between solid- (green), liquid- (yellow), and vapor-phase (red) molecules. Snapshots correspond to three distinct times, namely the moment of shock arrival at the trough of the surface asperity $t_{0}$, the time when the shock reaches the crest of the surface asperity $t_{0} + t_{\rm{c}}$, and time $t_{0} + 2t_{\rm{c}}$.

As expected, temperature increases with shock strength and the 2D temperature fields are generally quite similar between the small and large simulation cell cases. What is perhaps more interesting is that the corresponding zero-pressure phase near the surface asperity correlates strongly with key changes in jet morphology. At $u_{\rm{p}} = 0.5$~km/s, the temperature is always below the melting point of HMX and a jet never forms. In contrast, for the $u_{\rm{p}} = 1.0$~km/s cases where the jet first forms, the material in the shock focus region and jet reach temperatures above the melting temperature. Similarly, at $u_{\rm{p}} = 1.5$~km/s, more of the material near the surface exceeds the melting point and a very small fraction even exceeds the vaporization temperature in the 60~nm case. At and above $u_{\rm{p}} = 2.0$~km/s, the core of the jet exceeds the vaporization temperature, with the amount of vapor-phase molecules increasing with increasing $u_{\rm{p}}$. The cases with $u_{\rm{p}} \geq 2.0$~km/s are the only ones to exhibit the evolution of the incipient jet into a molecular plume. Clearly, a local transition to a fluid phase is highly correlated to jet formation, while transition to the vapor phase is seemingly necessary for that jet to expand into a plume.

\subsection{Asperity Wavelength Dependence}
\label{sec:ResultsWavelength}

The previous subsection considered the shock responses of an (010)-oriented HMX crystal surface asperity with a consistent aspect ratio ($kh_{0} = 3$) and two different asperity wavelengths ($\lambda = 30$~nm and 60~nm), which yielded qualitatively similar features. While morphology of the jet is similar in both cases for a given shock strength, notable differences in the jet head thickness can be seen in Figure \ref{fig:size}. Peak ejecta velocity $v_{z}^{99.9\%}$ is also modestly sensitive to the asperity wavelength, which was previously noted in connection with Table~\ref{tab:30nm-60nm-UsUp}. Here, we quantify differences between the two $\lambda$ cases in greater detail.

Spatially dependent field variables such as the particle velocity and temperature shown in Figures~\ref{fig:Up} and \ref{fig:size} can be concisely summarized as 1D distribution functions $f(x)$. For random variable $X$ of a given field $x$ with probability distribution $f(x)$, the 1D empirical cumulative distribution function (CDF)
\begin{equation}
  F_{X}(x) = \int_{-\infty}^{x} f_{X}(x')dx',
  \label{eq:cdf}
\end{equation}
and the empirical complementary cumulative distribution function (CCDF)
\begin{equation}
  \bar{F}_{X}(x) = 1 - F_{X}(x),
  \label{eq:ccdf}
\end{equation}
respectively quantify the probability that $X$ will take a value less than or greater than $x$. For this reason, the CCDF is often a more convenient and physically interpretable summary statistic than the underlying distribution function. For instance, the temperature CCDF evaluated at a given $T$ directly shows how much material is at or above temperature $T$. The CCDFs are computed at selected times over the entire simulation domain and are normalized by $N_{\rm{sd}}$, the number of molecules that fit in the sinusoidal asperity region defined in Figure~\ref{fig:system_cases}. Note that because each simulation contains $10N_{\rm{sd}}$ molecules, the normalized CCDF value can exceed unity.

\begin{figure}[t!]
  \centering
  \includegraphics[width=0.98\textwidth]{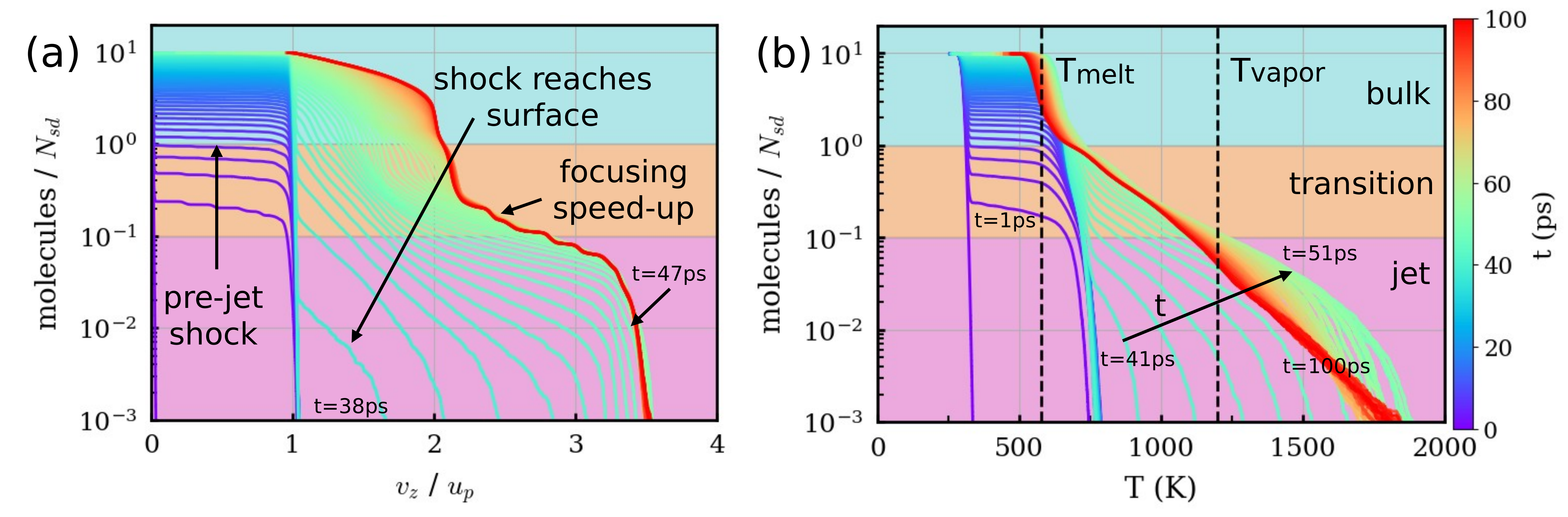}
  \caption{CCDFs of the (a) $v_{z}$ and (b) $T$ fields for the (010) case at $u_{\rm{p}} = 2.0$~km/s with surface asperity geometry factors $\lambda = 60$~nm and $kh_{0} = 3$. Each curve represents the CCDF taken at an instant in time with separation $\Delta t = 1$~ps, with curve color denoting the time of measurement. Blue, orange, and pink shaded areas denote empirically chosen bounds that qualitatively differentiate between the bulk, transition, and jet regions of the simulation.}
  \label{fig:CCDFs}
\end{figure}

Below we compute and compare the CCDFs for the $T$ and $v_{z}$ fields, but first we inspect the time dependence of these CCDFs for a baseline case. Figure~\ref{fig:CCDFs} shows the CCDFs of the $T$ and $v_{z}$ fields for the case with HMX (010) crystal orientation, surface asperity characteristics ($\lambda = 60$~nm, $kh_{0} = 3$), and shock strength $u_{\rm{p}} = 2.0$~km/s ($P_{zz} = 26$~GPa). We take this case as a baseline for comparison because the material jet behaves predominantly hydrodynamically, the molecular plume only partially develops, and the temperature field exhibits regions corresponding to solid, liquid, and vapor phases.

Focusing first on the $v_{z}$ field CCDF given in panel (a), one can clearly see the initial transit of the shock through bulk material, followed by initial surface interactions, and the eventual development of a steady-velocity jet. For times $t < 38$~ps, the shock transits the bulk and the material reaches a maximum velocity of $v_{z} = 1.0u_{\rm{p}}$. At $t = 38$~ps, the shock reaches the trough of the surface asperity and the material begins to freely expand into the vacuum. By $t = 40$~ps, shock focusing effects become clearly apparent as some molecules reach $v_{z} > 2.0u_{\rm{p}}$. Over the next 7~ps until $t = 47$~ps, the jet fully focuses and then reaches a steady-state velocity, which can be seen in the red curve, which overlaps curves at times 47~ps~$< t < 100$~ps. The CCDFs $\bar{F}_{X}$ facilitate qualitative categorization of molecules into distinct regions of the simulation. We take $\bar{F}_{X} > 10^{0}$ as defining the bulk region, $10^{-1} < \bar{F}_{X} \leq 10^{0}$ as defining a transition region, and $\bar{F}_{X} \leq 10^{-1}$ as defining the jet tip. For instance, only 10\% (or $0.1N_{\rm{sd}}$) of surface asperity molecules ultimately comprise the jet tip and reach the peak terminal velocity of $v_{z} \geq 3.0u_{\rm{p}}$.

The $T$ field CCDF in panel (b) exhibits a more complicated response compared to $v_{z}$. Here, for $t < 38$~ps, the shock transits through bulk material initially at 300~K, reaching a peak bulk shock temperature of nearly 750~K. The jet begins to form at $t = 38$~ps, but has a delayed temperature response until $t = 41$~ps, and does not reach its maximum temperature ($>$1800~K) until $t = 51$~ps. This delay is a few ps after the jet velocity reaches a steady state at $t = 47$~ps. From time $t = 43$~ps until the end of the simulation, the vapor-phase molecules in the jet expand and cool as they form a plume. This serves to both reduce the peak temperature of the jet (by roughly 50~K) and also its average temperature. For instance, while $0.01N_{\rm{sd}}$ molecules reach a peak temperature $>$1800~K at time $t = 51$~ps, most of these molecules cool to 1700~K by the end of the simulation at 100~ps. Additional plots of the $T$ field CCDFs for the $u_{\rm{p}} = 1.0$, 2.0, and 3.0~km/s cases are given in the SI.

Direct quantitative comparisons between the 30~nm and 60~nm cases can be made by extracting the time histories $X(t)$ of $\bar{F}_{X}$ at prescribed CCDF values that correspond to the average bulk ($\bar{F}_{X} = 5 \times 10^{0}$), transition ($\bar{F}_{X} = 5 \times 10^{-1}$), and jet ($\bar{F}_{X} = 5 \times 10^{-2}$) regions. For example, the time history of the average molecular velocity in the jet $v_{z}(t)$ is obtained parametrically by identifying the $v_{z}$ value for which $\bar{F}_{v_{z}} = 5 \times 10^{-2}$ at each point in time $t$. In what follows, the characteristic process timescale is twice as long for the (010)-60nm case as for the (010)-30nm case, so we normalize by the total simulation time $t_{\rm{end}}$, which is respectively 100~ps and 50~ps.

\begin{figure}[t!]
  \centering
  \includegraphics[width=0.98\textwidth]{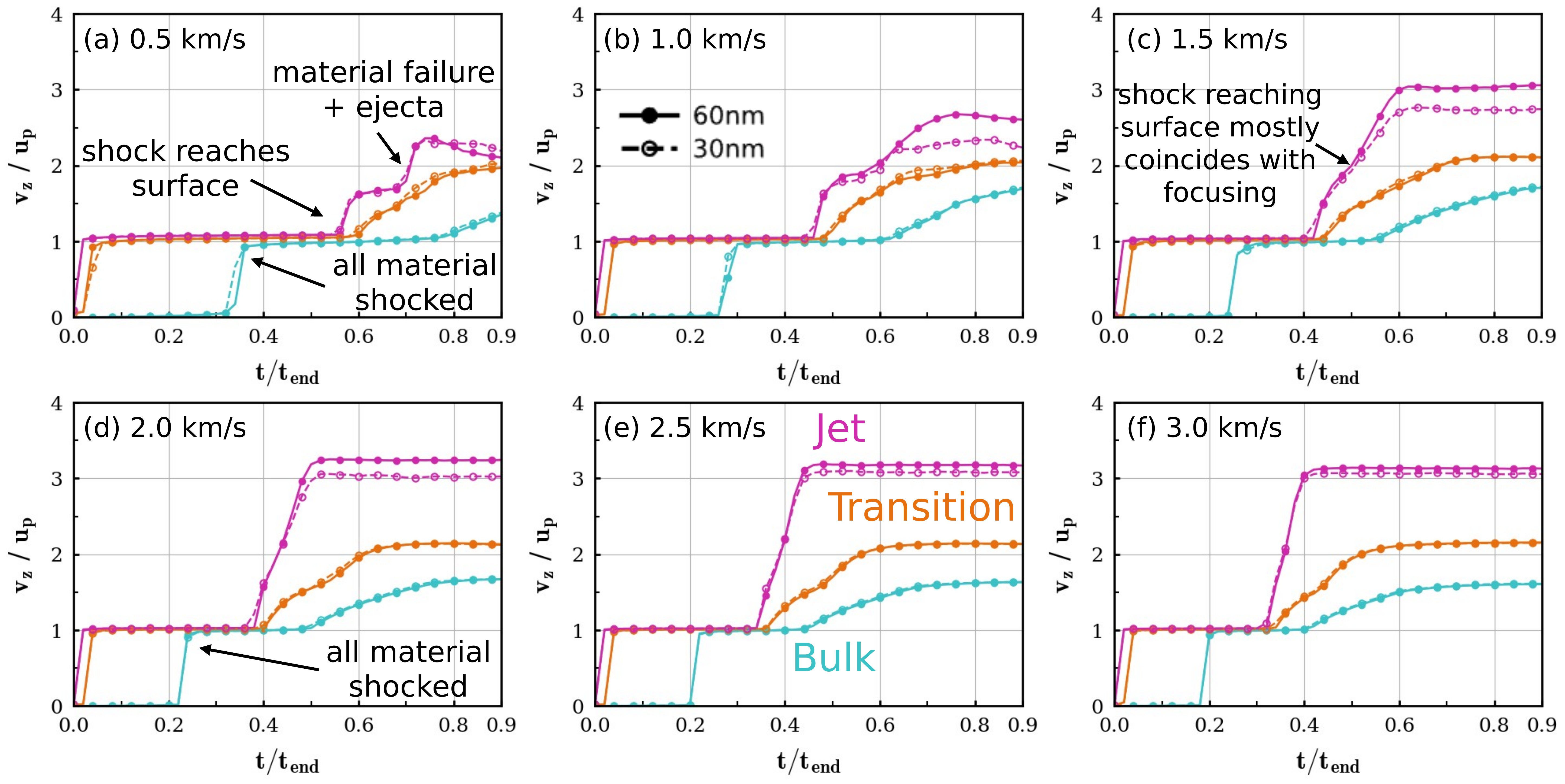}
  \caption{Comparison of the 30~nm (open symbols) and 60~nm (closed symbols) CCDF time histories $X(t)$ based on the $X = v_{z}$ CCDF at prescribed CCDF values corresponding to the bulk (cyan), transition (orange), and jet (magenta) regions shown for each shock strength case in panels (a)-(f). Note that time is scaled by $t_{\rm{end}}$, the total simulation time.}
  \label{fig:Up_comp}
\end{figure}

Figure~\ref{fig:Up_comp} shows comparisons of the $v_{z}(t)$ CCDF time histories for the 30~nm and 60~nm cases with each shock strength case shown in a separate panel. Several consistent trends hold across all $u_{\rm{p}}$ cases. First, the average molecular velocity in the bulk and transition regions is essentially identical between the 30~nm and 60~nm cases at all points in scaled time. Second, while the average jet velocity is similar for both sizes at each $u_{\rm{p}}$, when there is a deviation, it is almost exclusively the larger simulation that yields a higher $v_{z}(t)$. The $u_{\rm{p}} = 1.0$~km/s case exhibits the largest deviation in average jet velocity between the 30~nm and 60~nm cases, but these differences decrease with increasing $u_{\rm{p}}$ (a trend previously noted in connection with Table~\ref{tab:30nm-60nm-UsUp}). Third, material belonging to the transition region reaches $v_{z} = 2.0u_{\rm{p}}$, the value expected for free expansion of a shocked planar surface.

Comparing across the different $u_{\rm{p}}$ cases clearly shows a distinct transition in the response of the fastest molecules. The 0.5 and 1.0~km/s cases exhibit a two-step feature that corresponds to the initial shock interaction with the surface followed by a lagging step-wise focusing effect that drives up the peak velocity. As $u_{\rm{p}}$ increases, the time lag between the initial shock surface interaction and shock focusing decreases, producing a subtle inflection point in the 1.5~km/s case and a shock-focusing ramp for $u_{\rm{p}} \geq 2.0$~km/s. In terms of scaled $v_{z}(t)/u_{\rm{p}}$, there is very little quantitative difference in the time evolution of the three fastest $u_{\rm{p}}$ cases. This similarity in $v_{z}(t)$ is consistent with the morphological similarities of the jet's evolution into an expanding vapor plume seen previously for those three $u_{\rm{p}}$ cases in Figures~\ref{fig:Up} and \ref{fig:size}.

\begin{figure}[t!]
  \centering
  \includegraphics[width=0.98\textwidth]{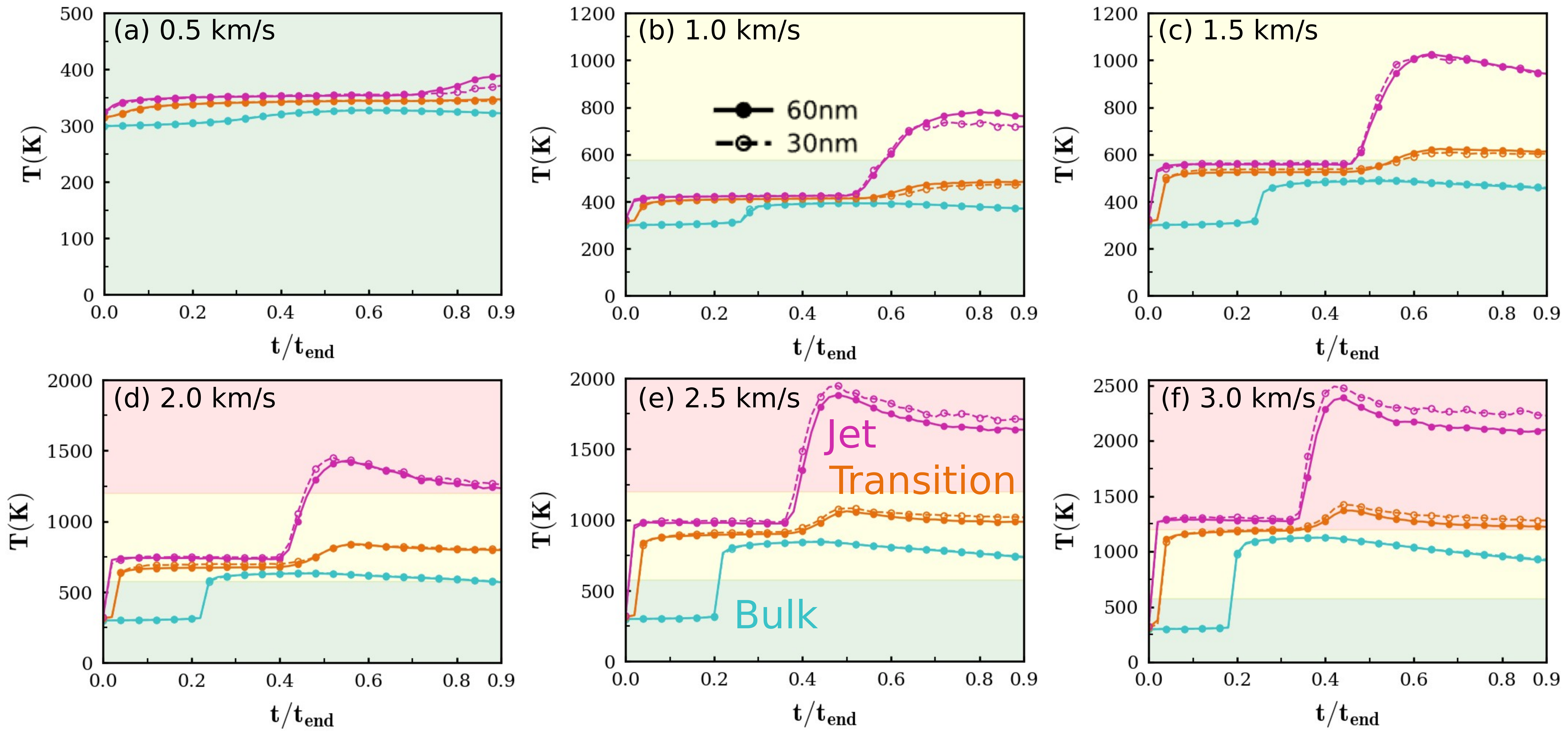}
  \caption{Comparison of the 30~nm (open symbols) and 60~nm (closed symbols) CCDF time histories $X(t)$ based on the $X = T$ CCDF at prescribed CCDF values corresponding to the bulk (cyan), transition (orange), and jet (magenta) regions shown for each shock strength case in panels (a)-(f). Note that time is scaled by $t_{\rm{end}}$, the total simulation time. Shaded panel background colors denote the solid (green), liquid (yellow), and vapor (red) regions of the phase diagram.}
  \label{fig:Temp_comp}
\end{figure}

Figure~\ref{fig:Temp_comp} shows analogous comparisons of the $T(t)$ CCDF time histories for the 30~nm and 60~nm cases with each shock strength case shown in a separate panel. Excursions into the liquid and vapor regions of the phase diagram are readily apparent in these plots. At $u_{\rm{p}} = 1.0$~km/s and 1.5~km/s, the jet enters the liquid phase, whereas for $u_{\rm{p}} \geq 2.0$~km/s the jet enters the vapor phase. At 1.5~km/s, the transition region is in the liquid phase while for $u_{\rm{p}} \geq 2.0$~km/s even the bulk reaches the liquid phase. These results hold important implications for microstructural-level responses in HMX under steady detonation conditions, for which the initial mechanical impulse of the detonation wave will be between $P_{\rm{CJ}}$ and $P_{\rm{VN}}$ ($u_{\rm{p}} = 3.0$~km/s being in this range). Under these conditions, rough microstructural free surfaces become nucleation points for the direct, localized vaporization of HMX into molecular plumes. Subsequent re-compression of these plumes is expected to be a highly efficient mechanism for hot spot formation.\cite{Holian2002}

Similar to $v_{z}(t)$, the $T(t)$ time histories are very similar between the 30~nm and 60~nm cases for each $u_{\rm{p}}$, with one distinct difference. Recall for $v_{z}(t)$, the bigger simulation size yielded the higher peak velocity. However, for $T(t)$, there is a distinct transition in size dependent differences with $u_{\rm{p}}$. For weak shocks, the bigger simulation reaches a higher temperature, but for strong shocks, it is the smaller simulation that gets hotter. We attribute this result to a confluence of two factors: (1) the effective strain rate, and (2) the rate-dependence of dissipation mechanisms in HMX. Two main dissipation mechanisms arise in these simulations, namely plastic deformation of the solid and viscous flow of the liquid. In shock-focusing geometries, it is well understood\cite{Wood2018, Rai2019, Kroonblawd2025} that the effective strain rate scales inversely with feature size,
\begin{equation}
    \dot\varepsilon \propto \frac{ \Delta u }{ \lambda },
    \label{eqn:strain-rate}
\end{equation}
where $\Delta u$ is the characteristic velocity difference across the deforming region. Thus, the 30~nm system experiences twice the effective strain rate of the 60~nm system.

Plastic deformation in crystalline HMX exhibits a rate- and pressure-hardening response,\cite{Zhang2023} while conventional strain hardening is comparatively weak.\cite{Khan2021} Similarly, the viscosity of amorphous HMX increases with shear rate.\cite{Khan2020} The higher characteristic strain rate in the 30~nm systems therefore can drive greater plastic dissipation, but rate-enhanced resistance to plastic flow also appears to hinder jet formation below $u_{\rm{p}} = 1.5$~km/s, lessening the contribution of focusing. For the two fastest $u_{\rm{p}}$, crystal strength plays a minimal role, since most surface material quickly reaches the liquid region of the phase diagram. Here, the rate-dependent viscosity implies greater dissipation in the smaller system, which is consistent with the greater degree of heating seen in those cases. These rate- and size-dependent responses seen in HMX are unlike the rate-independent strength responses of TATB crystal that result in effective scale invariance for shock-focusing geometries.\cite{Kroonblawd2025}

The molecular plumes that develop for strong shocks also exhibit size-dependent structural differences (recall Figure~\ref{fig:size}). We attribute plume formation to the fact that the jet core temperature exceeds the vaporization temperature at the pressure state of the jet. Both system sizes produce a plume at equivalent $u_{\rm{p}}$, but the larger system jet maintains a coherent outer surface layer for longer time duration, while the smaller system exhibits a more diffuse molecular plume. This difference may arise because the larger system contains more molecules and has a lower surface-to-volume ratio, leading to a physically larger jet with an initially liquid core. The characteristic evaporation timescale for a diffusion-limited process is expected to scale quadratically with droplet size. Thus, while the present systems exhibit a highly dynamic flow, it is reasonable to anticipate that physically larger jets would take longer to fragment and vaporize than smaller jets. It is also plausible that the smaller jet loses structure more quickly because of its greater proportion of surface molecules and encroachment on the spatial scale of molecular discreteness.

\subsection{Crystallographic Orientation Dependence}
\label{sec:ResultsOrientation}

\begin{figure}[t!]
  \centering
  \includegraphics[width=0.98\textwidth]{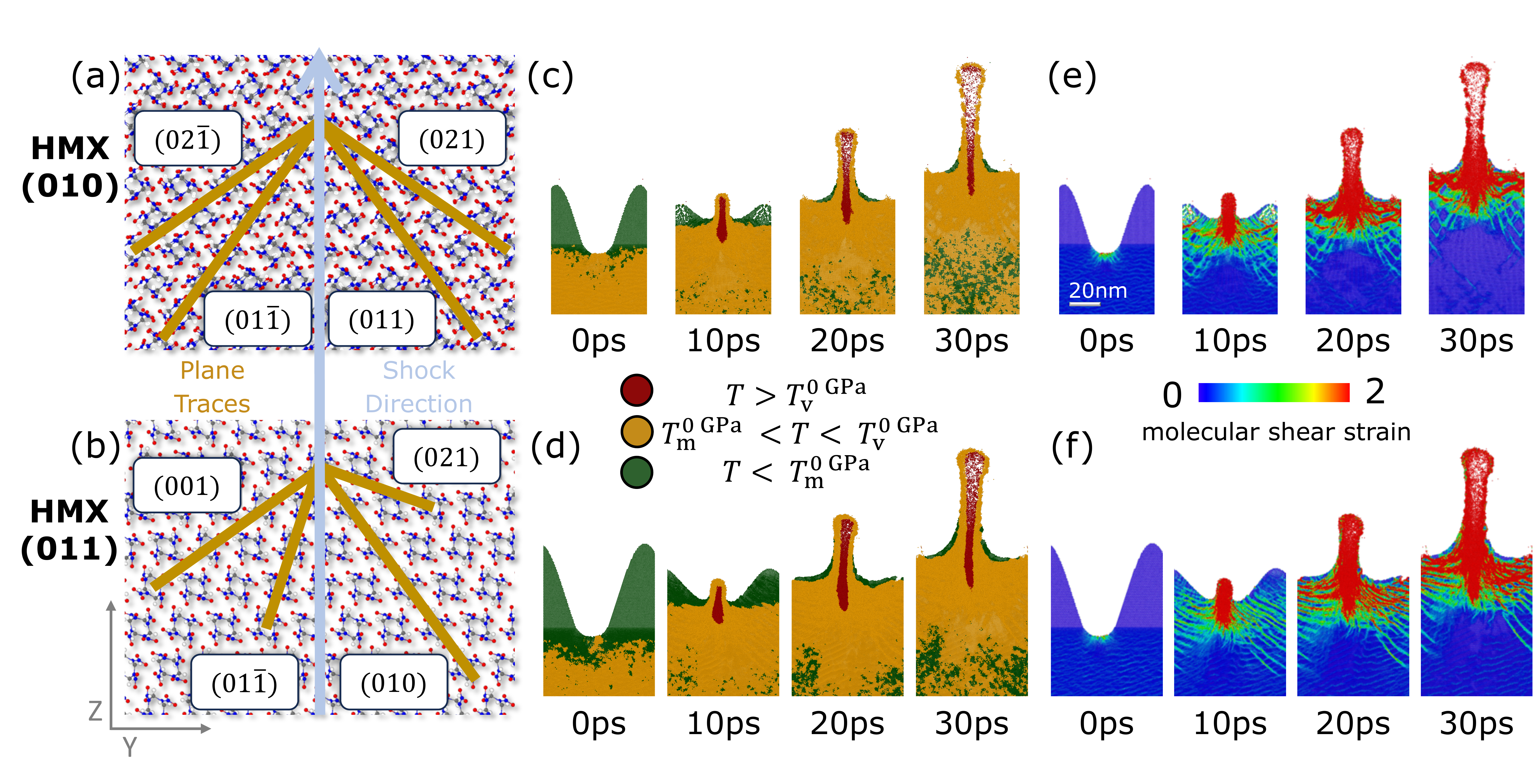}
  \caption{Comparison between the (010)-60nm case (top row) and the (011)-3 case (bottom row) with $kh_{0} \approx 3$ and $u_{\rm{p}} = 2.0$~km/s. (a,b) Crystal packing structure and slip plane orientation relative to shock direction. (c,d) Jet morphology snapshots at specified times with molecules colored according to a temperature-based phase assessment. (e,f) Analogous snapshots of the jet colored according to molecular shear strain.}
  \label{fig:orientation}
\end{figure}

Crystallographic orientation has been shown to affect shock-induced jetting at surface asperities in FCC metals.\cite{Li2014, Cherne2015} Unlike metals, HMX (like most organic explosives) crystallizes in a low-symmetry space group near standard conditions, which imparts strong directional dependencies in the elastic\cite{Mathew2018} and inelastic\cite{Zhang2023} mechanical response of HMX single crystals. For this reason, there is a strong likelihood for monoclinic ($P2_{1}/n$) $\beta$-HMX crystals to exhibit jetting characteristics that depend on shock orientation. To this end, we selected and studied jetting with shock directions oriented normal to the $\beta$-HMX (010) and (011) crystal facets. These two facets were chosen based on morphological importance\cite{Dick2004, Gallagher2015, Kroonblawd2016} and because they exhibit distinctly different symmetries about the shock axis. We focus here on a single representative comparison between the (010) and (011) responses for fixed $kh_{0} \approx 3$, shock strength $u_{\rm{p}} = 2.0$~km/s ($P_{zz} = 26$~GPa), and approximately equal asperity wavelength $\lambda_{(010)} = 60$~nm and $\lambda_{(011)} = 77$~nm. Note that $\lambda$ differs because the (011) crystal was prepared with future multi-component work in mind, which will study jet impacts on a down-stream lattice-matched crystalline target. Dependence on $kh_{0}$ will be examined in the following section. Figure~\ref{fig:orientation} shows snapshots of the molecular packing structure and jet morphology, which are discussed in turn.

Focusing first on underlying crystal packing in panels (a) and (b), it is clear that the (010) case exhibits a higher degree of symmetry about the shock axis $z$ in the plane of the quasi-2D simulation cell $y$-$z$. Both crystals are oriented such that $\mathbf{a} = [100]$ is along the $x$ direction (which is thin), intersections of slip planes with the $y$-$z$ plane are highlighted with yellow-gold lines, and a table of trace angles and interplanar spacings is given in the SI. As $\beta$-HMX belongs to the centrosymmetric point group $2/m$ with two-fold symmetry about $\mathbf{b} = [010]$, shock loading normal to the (010) facet leaves the loading configuration invariant under the symmetry operation $(0kl) \rightarrow (0k\bar{l})$. That is, there is mirror symmetry about the $x$-$z$ plane. In contrast, the (011) case has a facet normal that deviates by $54^\circ$ from the (010) facet normal, and more importantly lacks mirror symmetry about the shock axis. Above the Hugoniot elastic limit, inelastic deformation along the slip planes is expected, so the present two orientation cases test sensitivity of the jetting response to the symmetry of initial plastic flow in the crystal.

Inspection of the jet temperature/phase in panels (c) and (d) shows that jet symmetry is clearly correlated to underlying crystallographic symmetry. General features of both jets are similar. Each exhibits a vapor core and liquid outer shell. In the (010) case, the jet is oriented essentially exactly along $z$ and it exhibits structure with apparent pseudo-mirror symmetry apart from the seemingly stochastic breakup of the outer shell. In contrast, the (011) jet is oriented several degrees off-axis from $z$ and its liquid outer shell retains greater cohesion. From the snapshot series, it is clear that the off-axis flow in the (011) initiates at the earliest stages of material flow.

Recasting the jets in terms of molecular shear strain in panels (e) and (f) shows distinctly different plastic deformation responses in the two cases. In these examples, the shock pressure is $\approx$26~GPa. Dislocation mobility is hindered for many $\beta$-HMX slip systems at this pressure and shear localization is a predominant plastic deformation mechanism in this regime.\cite{Pal2018, Pal2019} The local shear strain mapping shows distinct shear bands that form near the surface and extend back into the bulk of the crystal. Qualitatively, there is some visual correlation between the shear band network geometry and the crystallographic slip planes in each case. In particular, the shear band network is decidedly asymmetric in the (011) case but exhibits greater mirror symmetry in the (010) case. We did not quantitatively associate shear bands to specific slip systems and also note that other slip planes may be relevant for the deformation response beyond those visible by looking along [100].

\begin{figure}[t!]
  \centering
  \includegraphics[width=3.35in]{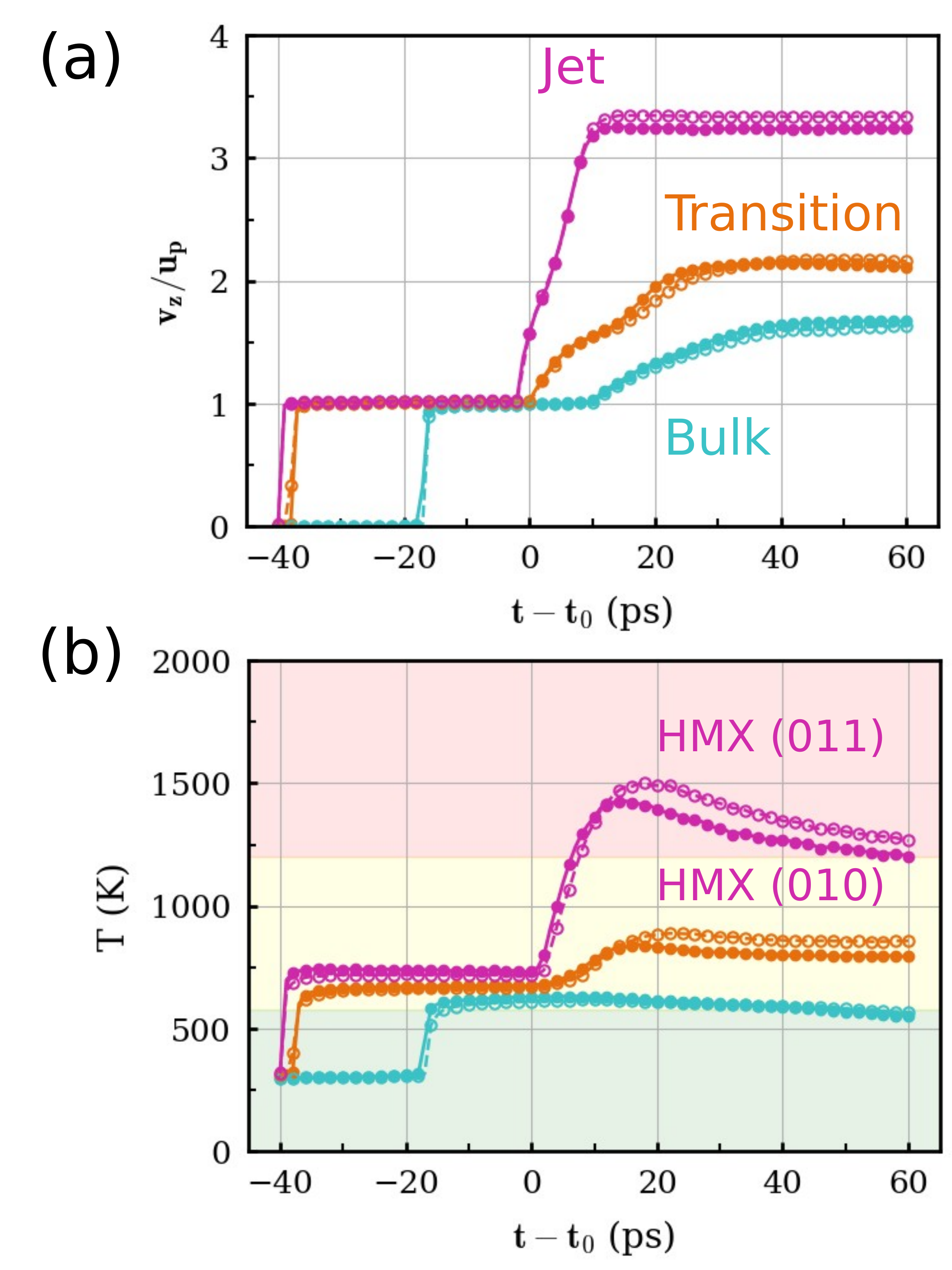}
  \caption{Time history of (a) the $v_{z}(t)$ and (b) $T(t)$ CCDFs evaluated in the bulk (cyan), transition (orange), and jet (magenta) regions with (010) and (011) cases plotted with closed and open symbols, respectively. CCDF region definitions are the same as in Figures~\ref{fig:Up_comp} and \ref{fig:Temp_comp}. Shaded panel background colors denote the solid (green), liquid (yellow), and vapor (red) regions of the phase diagram.}
  \label{fig:orientationCCDF}
\end{figure}

Figure~\ref{fig:orientationCCDF} shows CCDF time histories for $v_{z}$ and $T$, which demonstrates that the (010) and (011) systems exhibit quantitatively similar responses. Note that these plots are shown in physical (unscaled) time relative to $t_{0}$, the time at which the shock reaches the asperity trough, so the somewhat larger (011) case exhibits a small delay relative to the (010) case. The molecular velocity distribution is essentially identical in the bulk and transition regions, while the (011) jet tip reaches a marginally faster terminal velocity ($3.3u_{\rm{p}}$ compared to $3.2u_{\rm{p}}$). This difference could plausibly arise because the (011) cell $\lambda$ is 28\% larger than in the (010) cell, so it is unclear from the data at hand if this difference is a pure crystal orientation effect. Somewhat more substantial differences are apparent in the temperature field. Here, the (011) case seemingly undergoes more work leading to both a hotter jet tip and a hotter transition region. This difference is most likely due to differences in degree of initial crystal plastic deformation between the two cases, since the increased size of the (011) cell would lead to a slower strain rate and thus less viscous dissipation once the jet reaches the liquid state.

\subsection{Aspect Ratio Dependence}
\label{sec:ResultsAspectRatio}

\begin{figure}[t!]
  \centering
  \includegraphics[width=0.98\textwidth]{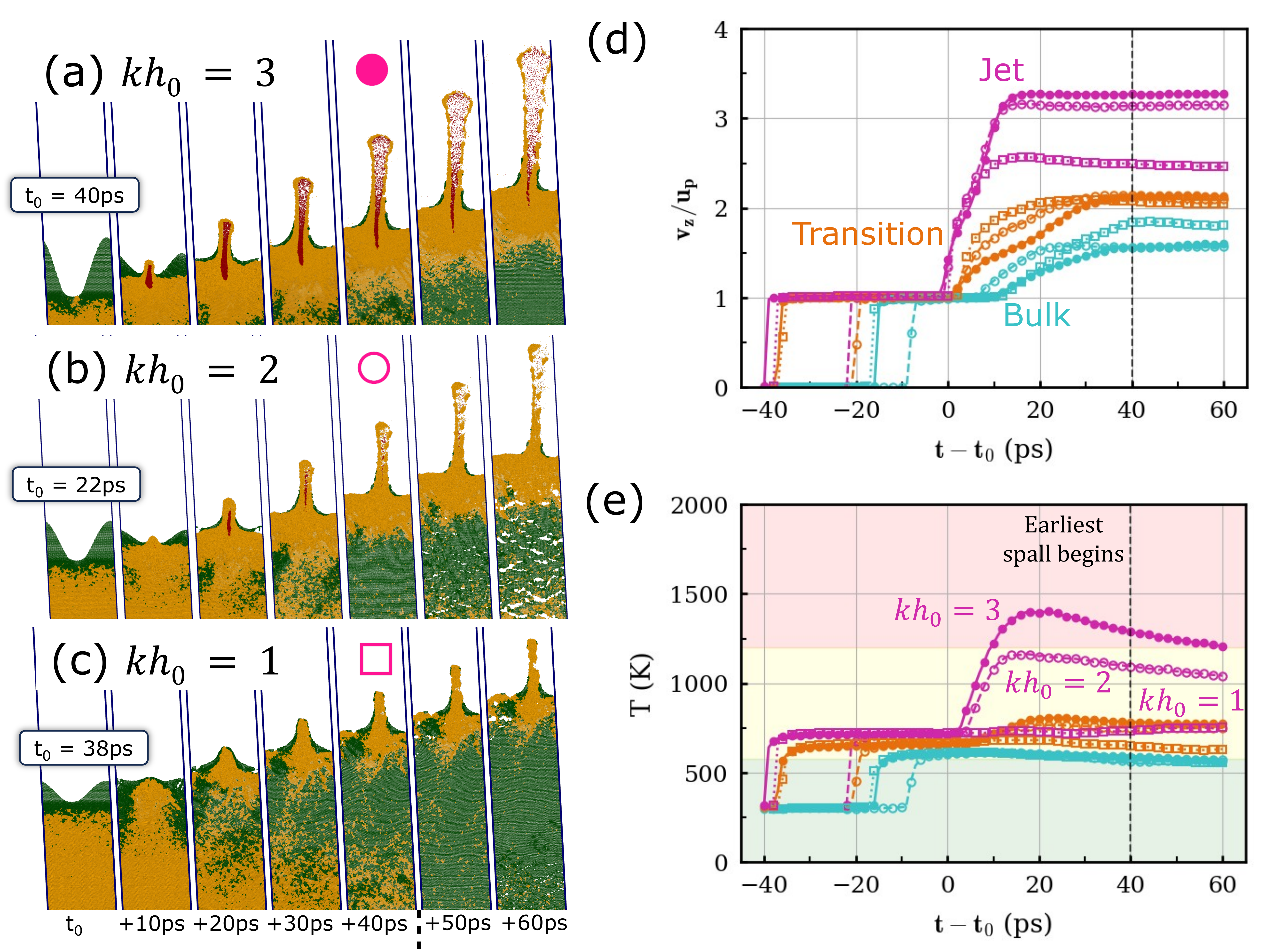}
  \caption{Simulation snapshots showing jetting response and temperature regimes for an (011)-oriented HMX crystal at  $u_{\rm{p}} = 2.0$~km/s with asperity geometry $\lambda = 77$nm and aspect ratios (a) $kh_{0} \approx 3$, (b) $kh_{0} \approx 2$, and (c) $kh_{0} \approx 1$. Molecule color conventions follow Figure~\ref{fig:size}. (d) Time histories of the $v_{z}$ CCDFs for the bulk (cyan), transition (orange), and jet (magenta) regions with symbols denoting cases $kh_{0} \approx 3$ (closed circles), $kh_{0} \approx 2$ (open circles), and $kh_{0} \approx 1$ (open squares). (e) Analogous time histories of the $T$ CCDFs with shaded color regions denoting the solid (green), liquid (yellow), and vapor (red) regions of the $P = 1$~atm HMX phase diagram.}
  \label{fig:aspect}
\end{figure}

Asperity aspect ratio is expected to strongly influence the thresholds for jet formation.\cite{Dimonte2010} All previously discussed simulations in this work considered an aspect ratio of $kh_{0} \approx 3$, which was expected (and shown) to be in the jetting regime for HMX with  $u_{\rm{p}} \geq 1.0$~km/s ($P_{zz} = 10$~GPa). However, it is unclear from these simulations alone whether there is a critical ratio below which shocked HMX asperities will fail to jet. To this end, we consider two additional simulations using the (011)-oriented HMX cell with $\lambda = 77$~nm and smaller sinusoid amplitudes $kh_{0} \approx 2$ and $kh_{0} \approx 1$ for comparison against the $kh_{0} \approx 3$ case. All three simulations set $u_{\rm{p}} = 2.0$~km/s ($P_{zz} = 26$~GPa), for which the $kh_{0} \approx 3$ case exhibited formation of a jet that evolved into a vapor plume. Figure~\ref{fig:aspect} shows a sequence of snapshots illustrating the jet formation process in each case along with time histories for the $v_{z}$ and $T$ CCDFs, which are discussed in turn.

Focusing first on the snapshot sequences in panels (a)-(c), it is clear that the asperity aspect ratio plays a strong role in governing the jet morphology. The bulk material also undergoes varying levels of spall, which is expected to be primarily dependent on sample length, since this controls the timing of the rarefaction wave reflection off the piston. For reasons owing to methodological consistency of ongoing work, the $kh_{0} = 2$ simulation had a shorter sample length of $L_{z} = 200~$nm $= 8h_{0}$. This leads to earlier spall in that case (beginning at $t_{0} + 40$~ps) relative to the other two $kh_{0}$ cases. These timing differences likely affect the late-time bulk response metrics, but should have little effect on the jet characteristics.

The baseline $kh_{0} \approx 3$ case exhibits a hot jet that expands into a large vapor plume, but the degree of heating decreases substantially with $kh_{0}$. The intermediate $kh_{0} \approx 2$ case exhibits only a very small fraction of molecules that reach the vaporization temperature. Although the resulting jet does expand, it breaks into distinct liquid droplets rather than forming a large vapor plume. The low-ratio $kh_{0} \approx 1$ case does not reach the vaporization temperature in any region. Here, a distinct liquid jet forms, but it retains a more cohesive structure that only begins to break apart by time $t_{0} + 60$~ps. The physical longitudinal dimension of the jet also decreases with decreasing $kh_{0}$. The jet morphology generated here with $kh_{0} \approx 1$ when $u_{\rm{p}} = 2.0$~km/s ($P_{zz} = 26$~GPa) is similar to the result for $kh_{0} \approx 3$ when $u_{\rm{p}} = 1.0$~km/s ($P_{zz} = 11$~GPa) seen previously in Figure~\ref{fig:Up}. While we do not extensively map the jetting threshold for HMX asperities as a function of $u_{\rm{p}}$ and $kh_{0}$, the propensity to form a jet clearly increases with increasing $u_{\rm{p}}$ and $kh_{0}$.

Inspection of the $v_{z}$ CCDFs in panel (d) reveals several distinct trends. First, the peak $v_{z}$ of the jet tip increases with increasing $kh_{0}$, but the response is nonlinear. Both the $kh_{0} \approx 3$ and $2$ cases reach very similar terminal velocities (respectively $3.3u_{\rm{p}}$ and $3.2u_{\rm{p}}$), but the $kh_{0} \approx 1$ case reaches a much smaller velocity ($2.5u_{\rm{p}}$). In contrast, the transition regions of each case reach a similar terminal velocity of $2.1u_{\rm{p}}$, where reaching the terminal velocity is likely dependent on trough-to-crest transit time. While the high-ratio cases exhibit faster jets, the low-ratio case exhibits a faster bulk velocity ($1.8u_{\rm{p}}$ vs. $1.5u_{\rm{p}}$). That is, as the aspect ratio decreases, the jet and bulk velocities converge towards $2.0u_{\rm{p}}$, the result expected for a planar surface. However, a momentum transfer trade-off in shock focusing is clear. Fast jet velocities arise at the expense of slow bulk velocities.

Heat generation during jet formation seen in panel (e) shows clear qualitative differences between the three aspect ratio cases. The baseline $kh_{0} \approx 3$ case jet exhibits a clear excursion past the vapor phase boundary, which is reflected in the expansion of the jet into a plume. In contrast, the jet formed in the two smaller aspect ratio cases stays within the liquid region of the phase diagram. The degree of localized heating is so low in the $kh_{0} \approx 1$ case that the jet formed in that case stays at or below the temperature of the transition region in the high-ratio $kh_{0} \approx 3$ case. All three cases exhibit a similar degree of bulk heating, with corresponding temperatures that are close to the solid-liquid phase boundary. Both $u_{\rm{p}}$ and $kh_{0}$ seemingly control the jet morphology through a similar mechanism. In both cases, the jet characteristics and its potential evolution into a plume are highly correlated with the local temperature relative to the solid-liquid and liquid-vapor phase boundaries. 

\subsection{Implications for Other Materials}

Across all cases considered, temperature presents as a consistent predictor of jet morphology. Jet formation seemingly occurs if the melting temperature is reached. Similarly, development of the jet into a molecular plume coincides with excursions beyond the vaporization temperature. Among the variables considered here, shock strength and aspect ratio exerted the strongest impacts on temperature and jet morphology. These results indicate that the ambient-pressure phase diagram may govern jetting morphology. For this reason, we anticipate that the thresholds for jet and plume formation will vary significantly across different materials.

Our study highlights a key difference in HMX as compared to the shock responses of metals, which are more extensively studied. Morphology of jets formed at surface asperities in metals maintain a more coherent and thin structure with a clear liquid-like bulb at the tip.\cite{Li2014, Cherne2015}. Modulation of asperity geometry has been explored in various studies on metals. While ejecta is common in those systems under certain conditions,\cite{Hamilton2025} the evolution of a jet into a molecular plume has not been observed in atomistic studies.\cite{Cherne2015, Li2014, Dyachkov2017, Hamilton2025} Based on the correlations between jet temperature, jet morphology, and ambient-pressure phase diagram identified here, we posit that jet morphology in metals may follow a similar underlying principle, but one that reflects a much larger difference between the 0~GPa melting and vaporization temperatures than seen here for HMX. For example, $T_{\rm{m}}^{\rm{Cu}} = 1358$~K and $T_{\rm{v}}^{\rm{Cu}} = 2835$~K ($\Delta T_{\rm{vm}}^{\rm{Cu}} = 1477$~K) and $T_{\rm{m}}^{\rm{Al}} = 933$~K and $T_{\rm{v}}^{\rm{Al}} = 2743$~K ($\Delta T_{\rm{vm}}^{\rm{Al}} = 1810$~K). These differences are more than twice the corresponding temperature range for HMX ($\Delta T_{\rm{vm}}^{\rm{HMX}} = 622$~K). MD simulations of very strong shocks in metals provide a direct route to test this hypothesis, with the understanding that the relevant phase barriers must be derived from the same interatomic potential used in the shock study.

Common organic explosives exhibit a wide range of melting temperatures at 0~GPa, with MD predictions yielding values of $T_{\rm{m}}^{\rm{TNT}} = 365$~K, $T_{\rm{m}}^{\rm{RDX}} = 482$~K, $T_{\rm{m}}^{\rm{HMX}} = 578$~K, and $T_{\rm{m}}^{\rm{TATB}} = 585$~K respectively for TNT,\cite{Holbrook2025} RDX,\cite{tow2022} HMX,\cite{tow2022} and TATB.\cite{Kroonblawd2024} Under the assumption of similar mechanical strength and local heating responses across these explosives, we anticipate that the shock pressure threshold for jet formation will rank similarly as the melting points for these materials. Similarly, the 0~GPa vaporization temperatures are respectively (and approximately) predicted by MD to be $T_{\rm{v}}^{\rm{TNT}} \leq 700$~K,\cite{Holbrook2025} $T_{\rm{v}}^{\rm{RDX}} \leq 1100$~K,\cite{Kroonblawd2022} $T_{\rm{v}}^{\rm{HMX}} \leq 1200$~K,\cite{Kroonblawd2023} and $T_{\rm{v}}^{\rm{TATB}} \leq 1250$~K,\cite{mathew_2018} although we note that these values have not been bounded as tightly as the melting points. The importance of accurately predicting highly coupled nonequilibrium thermomechanical processes such as jet formation provide strong motivations for thoroughly validating the MD FFs used in calculations. We do not explore FF sensitivity effects here for HMX, given that the Smith-Bharadwaj FF used in this work remains essentially the only high-quality nonreactive FF available for HMX. However, the explosive TATB has three well-validated nonreactive FFs, which both under- and over-predict the experimentally determined\cite{Bowlan2022} melting point (694~K) by 15-20\%;\cite{Kroonblawd2024} this makes TATB an interesting and practical case study for future FF sensitivity analyses. 

\section{Conclusions}
\label{sec:Conc}

Shock-induced jetting and recompression are plausible mechanisms for hot spot formation,\cite{Holian2002} but remain poorly understood in energetic molecular crystals. To this end, we performed MD simulations of shock interactions with a sinusoidal surface asperity in HMX crystals to examine the interplay between shock strength, asperity geometry, system size, and crystallographic orientation on jetting response. The shock pressures considered in this study span $\approx$4--47~GPa and are relevant for understanding HMX crystal responses under both detonation initiation and steady-state detonation conditions. The strongest shock was close to the von Neumann spike pressure, the value right after the shock front in a steady-state detonation. Across the cases examined here, the response evolves from strength-controlled viscoplastic deformation at low shock strength to coherent jetting and, under sufficiently strong focusing, to molecular plume formation. These transitions are governed not only by the shock pressure, but also by geometric effects impacting localized plastic dissipation, velocity amplification, and thermodynamic state of the ejecta.

Our simulations show that shock pressure and asperity aspect ratio play significant roles in determining shock response of HMX surface asperities. Several distinct transitions in material response were identified for a fixed asperity aspect ratio with high local curvature. The first of these is from a strength-controlled viscoplastic deformation regime at 4.5~GPa to onset of a jetting regime, which was observed near 11~GPa, the second-lowest shock pressure studied. At and above 26~GPa, shock focusing causes an increase in temperature of some HMX molecules to values above the ambient-pressure vaporization temperature, which is correlated with lateral jet expansion into a molecular plume. This effect is highly dependent on shock pressure, but additionally dependent on surface asperity curvature. Lower levels of curvature lead to less focusing, resulting in less plastic work and lower temperatures that hinder material reaching important phase boundaries for a given shock pressure.

Somewhat weaker functional dependencies were identified for asperity wavelength (i.e., physical size) and crystallographic orientation. Doubling the wavelength reduces the characteristic strain rate and changes the balance between plastic and viscous dissipation, which both exhibit rate-dependent resistance to deformation or flow.\cite{Zhang2023, Khan2021, Khan2020} For strong shocks, this produces greater heating in the smaller systems, whereas the size dependence is less uniform at lower shock strengths. Asperity wavelength also affects the stability of the liquid jet-tip shell, which is consistent with a size-dependent loss of molecular cohesion in the jet shell, although a critical diameter is not established here. Crystallographic orientation and symmetry of slip systems also play an important secondary role in jet formation. The primary influence of crystallographic orientation arises at the early stages of material flow, which can alter the trajectory angle of the resulting jet relative to the surface normal. Because our orientation comparisons also involved a modest difference in asperity wavelength, the orientation-specific contribution cannot be isolated completely.

Together, our findings show that jetting in molecular crystals exhibits distinct complexities that differ from those commonly reported in metals. For both HMX and metals, jetting response and ejecta morphology clearly depend on surface geometry, shock strength, and crystallographic orientation. In HMX, the coupling between low-symmetry molecular packing, crystal slip systems, phase-transition thresholds, and curvature focusing effects produces thicker, more fragmented ejecta and enables molecular plume formation under sufficiently strong shocks. As such, the initial microstructure geometry controls both the magnitude of localized heating but also the qualitative morphology of the ejecta. Formation of molecular plumes is seemingly connected to the limited stability range for HMX liquid. In contrast, metals typically exhibit a wider range of temperatures where the liquid remains stable. This suggests that plume-like expansion might also occur in metals under sufficiently strong shocks and high-curvature focusing, provided the local temperature crosses the corresponding liquid-vapor boundary. Impact and recompression of the simulated jets on downstream targets are the subject of ongoing work, which will help quantify the role of shock-induced jets in forming hot spots in HMX-based explosives. The high jet temperatures reached under some conditions also motivate ongoing reactive MD studies to quantify the coupling between phase changes and chemical decomposition during jet formation and upon impact with a downstream target.

\begin{acknowledgement}

This work was performed under the auspices of the U.S. Department of Energy by Lawrence Livermore National Laboratory under Contract DE-AC52-07NA27344. Work by Purdue University was supported by LLNL subcontract B648789. Approved for unlimited release, LLNL-JRNL-2026483-DRAFT.

The authors acknowledge RCAC and LLNL for computing resources.

\end{acknowledgement}


\begin{suppinfo}

Supplemental Information below.

\end{suppinfo}


\bibliography{bibtex}

\end{document}